\documentclass[conference]{IEEEtran}
\IEEEoverridecommandlockouts
\usepackage{cite}
\usepackage{amsmath,amssymb,amsfonts}
\usepackage{algorithmic}
\usepackage{graphicx}
\usepackage{textcomp}
\usepackage{xcolor}
\usepackage{array}
\usepackage{tabularx}
\usepackage{hyperref}
\usepackage{placeins}
\usepackage{float}

\hypersetup{
    colorlinks=true,
    linkcolor=blue,
    filecolor=magenta,      
    urlcolor=cyan,
    citecolor=green,
    pdftitle={Banking 2.0: The Stablecoin Banking Revolution},
    pdfauthor={Kevin McNamara, Rhea Pritham Marpu},
    pdfsubject={Financial Technology},
    pdfkeywords={stablecoins, banking, digital assets, financial technology, cryptocurrency},
    pdfproducer={LaTeX},
    pdfcreator={LaTeX}
}

\begin{document}

\title{Banking 2.0: The Stablecoin Banking Revolution\\
How Digital Assets Are Reshaping Global Finance}

\author{\IEEEauthorblockN{Kevin McNamara}
\IEEEauthorblockA{New Jersey, USA\\
kevin@mcnamara-group.com}
\and
\IEEEauthorblockN{Rhea Pritham Marpu}
\IEEEauthorblockA{New Jersey, USA\\
rm422@njit.edu}}

\maketitle

\begin{abstract}
The global financial system stands at an inflection point. Stablecoins represent the most significant evolution in banking since the abandonment of the gold standard, positioned to enable "Banking 2.0" by seamlessly integrating cryptocurrency innovation with traditional finance infrastructure. This transformation rivals artificial intelligence as the next major disruptor in the financial sector.

Modern fiat currencies derive value entirely from institutional trust rather than physical backing, creating vulnerabilities that stablecoins address through enhanced stability, reduced fraud risk, and unified global transactions that transcend national boundaries.

Recent developments demonstrate accelerating institutional adoption: landmark U.S. legislation including the GENIUS Act of 2025, strategic industry pivots from major players like JPMorgan's crypto-backed loan initiatives, and PayPal's comprehensive "Pay with Crypto" service. Widespread stablecoin implementation addresses critical macroeconomic imbalances, particularly the inflation-productivity gap plaguing modern monetary systems, through more robust and diversified backing mechanisms.

Furthermore, stablecoins facilitate deregulation and efficiency gains, paving the way for a more interconnected international financial system. This whitepaper comprehensively explores how stablecoins are poised to reshape banking, supported by real-world examples, current market data, and analysis of their transformative potential.
\end{abstract}


\section{Introduction}

\subsection{The Great Financial Transformation}

The global banking landscape is experiencing its most profound transformation since the Bretton Woods system collapsed in 1971. We are rapidly transitioning from a system heavily dependent on physical assets and centralized intermediaries toward one increasingly dominated by virtual currencies and decentralized infrastructure.

This shift fundamentally reimagines how value is stored, transferred, and managed globally. Our thesis posits that stablecoins will become the foundational infrastructure of future banking systems \cite{boker2023explainable}, offering a stable alternative that unifies global transactions, reduces fees and settlement times \cite{bis2018cbdc}, and delivers superior value to end-users amid accelerating cryptocurrency adoption worldwide \cite{sergio2025global}.

This whitepaper will examine the historical context of stablecoins, their distinct advantages over traditional banking systems, the challenges they face, recent pivotal developments including legislative actions and corporate adoptions, and their transformative future outlook. Our methodology involves rigorous analysis based on established economic principles, recent regulatory changes, and illustrative case studies drawn from current industry insights and reliable data sources.

\begin{figure*}[t]
    \centering
    \includegraphics[width=0.8\textwidth]{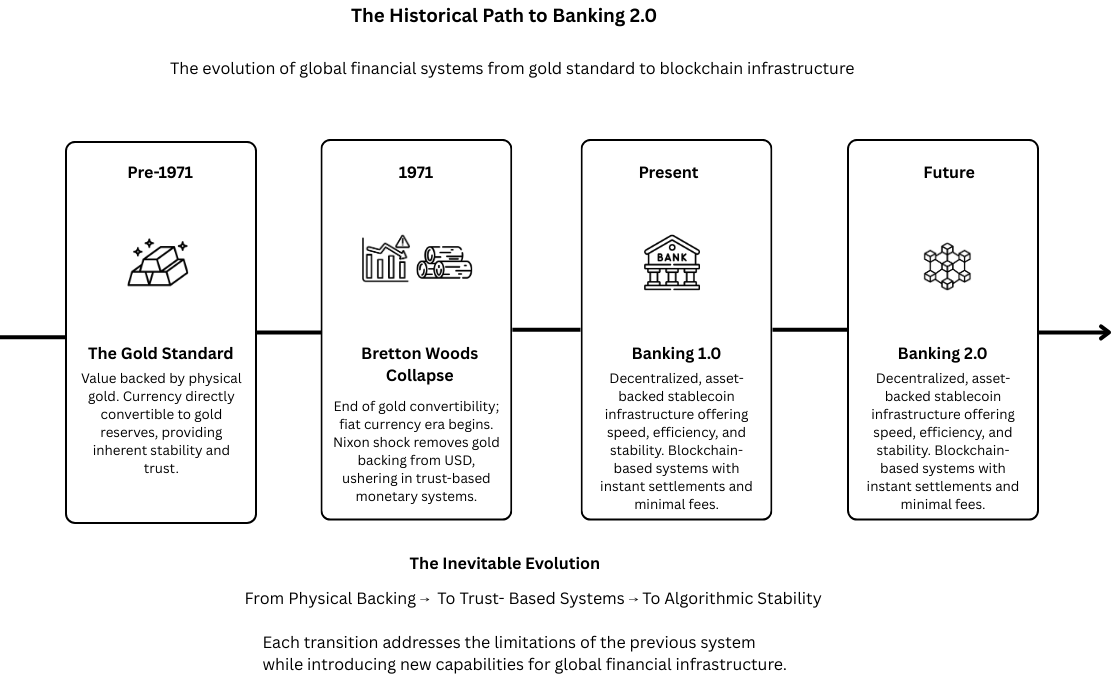}
    \caption{Timeline of Evolution to Banking 2.0}
    \label{fig:historical_timeline}
\end{figure*}

Figure~\ref{fig:historical_timeline} illustrates the key milestones in the evolution from traditional banking to the emerging Banking 2.0 paradigm, highlighting critical developments in monetary policy, technological innovation, and regulatory frameworks.

\section{The Nature of Modern Currencies and Monetary Systems}

\subsection{The Virtual Reality of Contemporary Money}

A fundamental understanding of modern monetary systems reveals that nearly all major currencies are inherently virtual due to the practice of unlimited monetary expansion by central banks \cite{fedreserve1971}. Since the abandonment of the gold standard, currency values have become functions of economic stability and market confidence rather than tangible asset backing. This represents a dramatic shift from historical precedent, where currencies were typically backed by precious metals or other physical commodities.

The implications of this virtual nature are profound. When currencies can be created with keystrokes rather than mined from the earth or manufactured through productive economic activity, their stability becomes entirely dependent on institutional credibility and economic performance. This system works adequately during periods of economic growth and political stability but becomes vulnerable during crises, as evidenced by various currency collapses throughout modern history.

Notable exceptions exist in countries that have strategically maintained stronger backing mechanisms. Switzerland, Singapore, the United Arab Emirates, and Saudi Arabia have preserved substantial physical gold reserves while increasingly exploring crypto reserves as additional backing \cite{renteria2021bitcoin}. The UAE exemplifies this trend, having solidified its status as a burgeoning crypto hub by facilitating more than \$300 billion in regional crypto transactions \cite{reuters2025abu} while simultaneously boosting its gold reserves by 19.3\% in the first quarter of 2025 \cite{wam2025cbuae}. This dual approach -- maintaining traditional safe-haven assets while embracing digital alternatives -- represents a sophisticated hedging strategy against monetary instability \cite{chainalysis2023bitcoin}.

\subsection{The Inflation-Productivity Imbalance Crisis}

The long-term viability of global monetary systems critically depends on maintaining a healthy balance between inflation and productivity growth. Currently, many systems exhibit an alarming characteristic: algorithmically inflated money supplies coupled with stagnant or declining productivity growth \cite{cea2025economic}. This imbalance creates a structural problem that inevitably leads to currency devaluation over time.

The United States provides a stark example of this phenomenon. Despite technological advances and efficiency gains in various sectors, overall productivity growth has remained relatively flat while the money supply has expanded dramatically, particularly following the 2008 financial crisis and the COVID-19 pandemic response. This divergence creates persistent inflationary pressure that erodes purchasing power and savings, disproportionately affecting middle- and lower-income populations \cite{cea2025economic}.

Traditional monetary policy tools -- primarily interest rate adjustments -- have proven insufficient to address this structural imbalance. Central banks find themselves trapped between the need to combat inflation and the political imperative to maintain economic growth through monetary stimulus. This policy paradox creates systemic instability that stablecoins, with their alternative backing mechanisms, are positioned to help resolve.

\subsection{Inherent Limitations of Fiat Systems}

While fiat currencies derive legitimacy from their respective national economies, they remain inherently susceptible to volatility, fraud, and political manipulation within centralized systems. This vulnerability extends to traditional financial infrastructure, where single points of failure, bureaucratic inefficiencies, and rent-seeking intermediaries create friction and risk throughout the system.

The centralized nature of fiat systems creates multiple attack vectors for fraud and manipulation. From currency counterfeiting to electronic payment fraud, traditional systems require extensive and expensive security infrastructure that ultimately raises costs for all participants. Moreover, the opacity of many traditional financial transactions makes it difficult to detect and prevent fraudulent activity until significant damage has occurred.

These systemic vulnerabilities highlight the need for alternative monetary systems that can provide stability without inheriting the structural weaknesses of traditional fiat currencies \cite{fedreserve1971}, \cite{cea2025economic}.

Table~\ref{tab:banking_evolution} provides a comprehensive comparison of traditional fiat systems versus stablecoin-based infrastructure, highlighting the fundamental differences between Banking 1.0 and Banking 2.0 architectures.

\begin{table}[htbp]
    \caption{The Evolution from Banking 1.0 to Banking 2.0}
    \label{tab:banking_evolution}
    \centering
    \small
    \begin{tabular}{|p{2cm}|p{2.6cm}|p{2.6cm}|}
    \hline
    \textbf{Feature} & \textbf{Banking 1.0} & \textbf{Banking 2.0} \\
    \hline
    Foundation & Centralized Intermediaries(Bank \& SWIFT) & Decentralized Blockchain Networks \\
    \hline
    Value Backing & Institutional Trust \& National Economies & Diversified Reserves (Fiat, Assets, Crypto) \\
    \hline
    Cross Border Speed & 2-5 Business Days & Near Instant Minutes \\
    \hline
    Transaction Fees & High, \% Based & Low, Fixed \\
    \hline
    Security Model & Centralized, Single Points of Failure & Decentralized, Cryptographically secured \\
    \hline
    Transparency & Opaque(Internal Ledgers) & Transparent (Public Ledger) \\
    \hline
    Key Weakness & Inflation, Friction, Counterparty Risk & Peg Stability, Regulatory Fragmentation \\
    \hline
    \end{tabular}
    \begin{flushleft}
    \small
    \textit{Banking 1.0: Traditional fiat systems. Banking 2.0: Stablecoin systems.}
    \end{flushleft}
\end{table}

\section{The Rise of Stablecoins: History and Evolution}

\subsection{A Decade of Real-World Testing}

Contrary to popular perception, stablecoins are not a recent innovation. They have been operational and subjected to rigorous real-world testing in open, decentralized environments for more than a decade \cite{boker2023explainable}. Early implementations such as BitUSD and NuBits emerged around 2014, providing valuable lessons about the challenges and opportunities of maintaining price stability in digital assets \cite{boker2023explainable}.

This extended testing period represents an unprecedented stress test for any financial innovation. Unlike traditional financial products that are developed in controlled laboratory environments and gradually rolled out, stablecoins have been subjected to constant scrutiny from a global community of developers, security experts, and adversarial actors. This exposure has included sophisticated hacking attempts, market manipulation efforts, and various forms of technological assault.

The transparency of blockchain transactions enables real-time monitoring that can detect suspicious patterns immediately. While traditional financial crime often goes undetected for months or years, blockchain-based fraudulent activity becomes apparent within minutes, allowing for rapid response and mitigation \cite{benos2019economics}. This transparency makes illicit activities far more traceable than in traditional systems, where transactions can be obscured through complex intermediary networks and proprietary processing systems \cite{benos2019economics}.

\subsection{Evolutionary Mechanisms and Types}

The stablecoin ecosystem has evolved to encompass several distinct approaches to maintaining price stability. Fiat-pegged stablecoins, such as USDC and USDT, represent the most straightforward approach, attempting to maintain a one-to-one peg with specific national currencies. These instruments serve as bridges between traditional and digital finance, enabling users to maintain exposure to familiar value references while benefiting from blockchain efficiency.

However, the evolution of stablecoin design has revealed more sophisticated approaches. Algorithmic stablecoins attempt to maintain stability through programmatic supply adjustments, while asset-backed stablecoins derive their stability from diversified reserve portfolios. Each approach offers distinct advantages and faces unique challenges, contributing to a rich ecosystem of experimental monetary instruments.

For true global financial stability, there is growing recognition of the need for internationally available digital currencies whose value derives from the intrinsic worth of underlying infrastructure as a public commodity, rather than solely from volatile fiat pegs. This concept echoes the historical role of gold in underpinning currency values, suggesting a future where stablecoins are backed by diversified, resilient reserves including precious metals, real estate, productive assets, and other cryptocurrencies.

\subsection{Transformative Impact on Transaction Infrastructure}

Stablecoins fundamentally reshape how transactions occur by unifying them under a single, virtual pipeline that eliminates numerous traditional intermediaries. Traditional cross-border payments typically involve multiple banks, clearinghouses, correspondent banking relationships, and regulatory checkpoints, each adding fees, delays, and potential failure points \cite{bis2018cbdc}.

By contrast, stablecoin transactions can occur directly between parties through blockchain networks, settling in minutes rather than days and incurring fees measured in cents rather than dollars or percentages \cite{bis2018cbdc}. This efficiency gain is not merely incremental -- it represents an order-of-magnitude improvement in transaction economics that makes previously uneconomical use cases viable.

The impact extends beyond individual transactions to the broader structure of global commerce. When transactions can occur seamlessly across borders without the friction of currency conversion, regulatory arbitrage, and intermediary fees, it enables new forms of economic organization and collaboration. Small businesses can access global markets more easily, developing economies can participate more fully in international trade \cite{sergio2025global}, and individuals can preserve value across unstable monetary regimes.

\begin{figure*}[t]
    \centering
    \includegraphics[width=\textwidth]{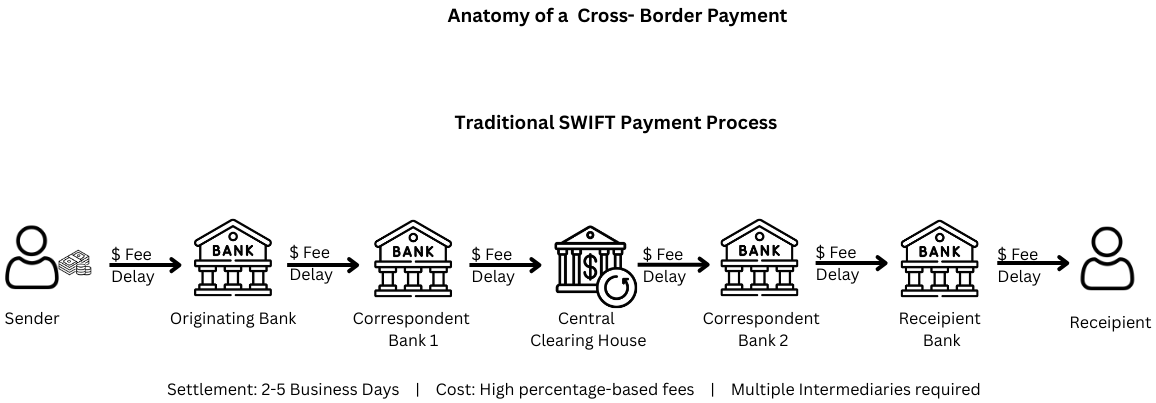}
    \caption{Traditional cross border Payment}
    \label{fig:traditional_payment}
\end{figure*}

Figure~\ref{fig:traditional_payment} demonstrates the complex multi-step process involved in traditional cross-border payments, illustrating the numerous intermediaries and potential failure points that stablecoins can eliminate.

\section{Advantages of Stablecoins in Banking 2.0}

\subsection{Superior Fraud Resistance and Security Architecture}

The open and transparent nature of blockchain technology underlying stablecoins has subjected them to continuous scrutiny and testing by a global community of security experts, white-hat hackers, and adversarial actors. This extensive exposure has paradoxically made them more secure and fraud-resistant than traditional fiat systems, which rely on centralized security measures and remain vulnerable to internal fraud and single points of failure \cite{benos2019economics}.

Traditional banking systems concentrate risk in centralized databases and processing centers that, while heavily guarded, present attractive targets for sophisticated attackers. The 2017 Equifax breach, various SWIFT network attacks, and countless payment processor compromises demonstrate the vulnerability of centralized financial infrastructure. When these systems are compromised, damage can be extensive and difficult to detect until significant losses occur.

Blockchain-based stablecoins distribute security across thousands of nodes worldwide, making successful attacks exponentially more difficult. Every transaction is cryptographically secured and verified by multiple independent parties before being permanently recorded. This architecture creates a security model that improves with scale rather than becoming more vulnerable as adoption increases.

The transparency of blockchain transactions enables real-time monitoring that can detect suspicious patterns immediately \cite{benos2019economics}. While traditional financial crime often goes undetected for months or years, blockchain-based fraudulent activity becomes apparent within minutes, allowing for rapid response and mitigation.

\subsection{Revolutionary Global Transaction Efficiency}

The inherently decentralized nature of stablecoins enables significant deregulation opportunities that can lead to greater user control, dramatically lower fees, and unparalleled ease for international financial transfers. This potential is being realized through initiatives like the Global Dollar Network (GDN), which involves prominent firms including Robinhood, Kraken, and Paxos working to accelerate stablecoin adoption for global payments \cite{gdn2025global}, \cite{mcclain2024introducing}.

These networks challenge established incumbents like traditional SWIFT-based international transfers by sharing revenues among participants and streamlining cross-border transactions. The result is a more competitive and efficient marketplace where costs are driven down through technological innovation rather than regulatory protection of incumbent monopolies.

The efficiency gains extend beyond cost reduction to fundamental improvements in settlement speed and reliability. While traditional international transfers can take days to complete and may fail for various bureaucratic reasons, stablecoin transfers typically settle within minutes with cryptographic certainty \cite{bis2018cbdc}. This improvement in speed and reliability enables new business models and economic relationships that were previously impractical.

\begin{figure}[t]
    \centering
    \includegraphics[width=\columnwidth]{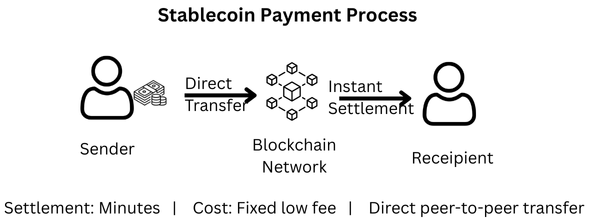}
    \caption{Stablecoin cross border Payment}
    \label{fig:stablecoin_payment}
\end{figure}

Figure~\ref{fig:stablecoin_payment} contrasts the streamlined stablecoin payment process with traditional methods, showcasing the dramatic reduction in intermediaries and settlement time.

\subsection{Addressing Macroeconomic Stability Challenges}

Stablecoins hold considerable potential to help balance the inflation-productivity equation by offering a more stable store of value that is less susceptible to inflationary pressures from unlimited currency printing \cite{cea2025economic}. This capability becomes increasingly important as traditional monetary policy tools prove inadequate for addressing structural economic imbalances.

Advanced stablecoin implementations like Ripple's RLUSD, which integrates with Chainlink's decentralized oracle network, are designed to enhance liquidity and accessibility within DeFi and traditional banking ecosystems \cite{doe2025ripple}. These systems can potentially draw liquidity away from purely fiat-backed alternatives by providing enhanced transparency, compliance capabilities, and stability mechanisms that don't rely solely on central bank credibility.

The macroeconomic implications are profound. If significant portions of global commerce migrate to stablecoin-based systems, it could reduce the ability of individual nations to export inflation through monetary expansion. This constraint could force more responsible fiscal policies while providing individuals and businesses with more stable stores of value that aren't subject to political manipulation.

\subsection{Seamless Integration with Traditional Finance}

The growing acceptance and integration of stablecoins by traditional financial institutions marks a significant inflection point in the evolution toward Banking 2.0. Pioneering institutions like Kraken Bank in Wyoming have established new models for bridging crypto and fiat currencies while maintaining regulatory compliance and consumer protection \cite{kraken2020wyoming}.

Operating as a Special Purpose Depository Institution with 100\% reserves, Kraken Bank offers services including crypto custody and fiat deposit accounts, setting precedents for regulated and secure digital asset banking \cite{kraken2020wyoming}. This model demonstrates that stablecoin integration doesn't require abandoning regulatory oversight or consumer protections -- it simply requires updating regulatory frameworks to accommodate technological innovation.

The integration process is accelerating as traditional financial institutions recognize both the competitive threat and the opportunity presented by stablecoins. Rather than being displaced by this technology, forward-thinking institutions are positioning themselves to benefit from the efficiency gains and new revenue opportunities that stablecoin integration provides.

\section{Recent Developments and Case Studies}

\subsection{Landmark U.S. Legislation: The GENIUS Act of 2025}

On July 18, 2025, President Donald Trump signed the Guiding and Establishing National Innovation for U.S. Stablecoins Act (GENIUS Act) into law, establishing the first comprehensive federal framework for U.S. dollar-pegged stablecoins in the United States \cite{whitehouse2025genius}. This landmark legislation represents a fundamental shift in regulatory approach, moving from prohibition and restriction toward enabling innovation within appropriate guardrails.

The Act mandates that qualifying stablecoins maintain 100\% reserves in highly liquid assets, predominantly U.S. Treasuries, creating a new category of digital assets that combine the efficiency of blockchain technology with the stability of government-backed securities \cite{housefinance2025genius}. This requirement addresses one of the primary concerns about early stablecoin implementations -- the quality and transparency of reserve assets.

The market impact was immediate and significant. Crypto-related Exchange Traded Funds (ETFs) experienced substantial inflows following the Act's passage, signaling increased institutional investor confidence and validation of the regulatory clarity provided \cite{chauhan2025crypto}. This response demonstrates how appropriate regulation can catalyze rather than constrain innovation and adoption.

The legislation also establishes clear operational requirements for stablecoin issuers, including regular auditing, consumer protection measures, and interoperability standards \cite{housefinance2025genius}. These provisions create a framework that protects consumers while enabling innovation, providing a model that other jurisdictions are likely to emulate.

\subsection{Strategic National Asset Management}

President Trump's strategic decision to establish a sovereign wealth fund that could potentially include Bitcoin and stablecoins as national reserves represents a revolutionary approach to government asset management. This policy aims to leverage projected growth in crypto valuations to reduce national debt, representing a novel approach that treats digital assets as strategic national resources rather than speculative investments \cite{labonte2025banking}.

The formation of a crypto advisory council, reportedly including prominent industry figures such as Ripple's CEO, demonstrates institutional commitment to developing sophisticated digital asset strategies \cite{whitehouse2025digital}. This council provides a mechanism for ongoing policy development and ensures that government initiatives remain aligned with technological and market developments.

The broader implications extend beyond national finance to international monetary relations. If the United States successfully integrates digital assets into its strategic reserves, other nations will likely follow, potentially accelerating the global transition toward stablecoin-based international commerce \cite{chainalysis2023bitcoin}.

\subsection{Corporate Transformation: JPMorgan's Strategic Pivot}

JPMorgan Chase's evaluation of crypto-backed loan offerings represents one of the most significant corporate reversals in the financial services industry \cite{smith2025jpmorgan}. Under CEO Jamie Dimon's leadership, the bank has historically maintained a cautious, often critical stance toward cryptocurrencies, making this potential shift particularly noteworthy.

The reported plans to offer crypto-backed loans using Bitcoin and Ethereum as collateral, with a potential launch by 2026, signal a broader recalibration within the traditional banking sector, recognizing the undeniable demand and utility of digital assets \cite{smith2025jpmorgan}.

The development is particularly significant because JPMorgan's institutional credibility and regulatory relationships provide a model for other traditional banks to follow. When the largest bank in the United States begins treating cryptocurrencies as legitimate collateral assets, it removes significant barriers to broader institutional adoption.

This corporate evolution also demonstrates the practical maturation of cryptocurrency markets. The ability to use digital assets as loan collateral requires sophisticated risk management frameworks, regulatory clarity, and market stability -- all of which are increasingly available in the current environment.

\subsection{PayPal's Comprehensive Crypto Integration}

PayPal's launch of its "Pay with Crypto" service represents a watershed moment in mainstream cryptocurrency adoption. This initiative allows merchants to accept dozens of cryptocurrencies, automatically converted into PayPal's native PYUSD stablecoin, creating a seamless bridge between digital assets and traditional commerce.

The service offers compelling economic advantages with transaction fees starting at 0.99\% for the first year before increasing to 1.5\%--notably lower than typical international credit card processing fees. Merchants benefit from near-instant access to proceeds and a 4\% yield on PYUSD holdings, providing operational efficiency improvements and additional revenue incentives.

The international implications are transformative. PayPal envisions cross-border commerce becoming as simple as domestic transactions, addressing persistent pain points in global payment complexity and costs. This capability enables seamless transactions between any global participants without traditional banking intermediaries.

With PYUSD achieving approximately \$900 million market capitalization as the 12th-largest stablecoin, PayPal has demonstrated traditional payment processors' potential to successfully launch and scale stablecoin offerings, providing a roadmap for broader financial sector integration.

\subsection{Global Innovation and Market Development}

Beyond U.S. developments, international initiatives are rapidly advancing stablecoin adoption and integration. The UAE continues expanding its role as a significant DeFi hub, with companies like DeFi Technologies leading quantum-secure stablecoin network development \cite{nist2022quantum}. This focus on quantum security addresses one of the most significant long-term threats to blockchain-based financial systems.

Singapore plays a crucial role in facilitating stablecoin issuance and global adoption through its participation in networks like the Global Dollar Network \cite{gdn2025global}. The country's sophisticated regulatory framework and strategic position in Asian markets make it an ideal hub for stablecoin innovation and distribution.

Companies such as Kraken and Robinhood are actively developing stablecoin network infrastructure that facilitates global payments while enhancing accessibility for retail and institutional users \cite{gdn2025global}, \cite{mcclain2024introducing}. These efforts create competitive alternatives to established payment networks while driving down costs and improving service quality.

\subsection{Market Validation Through ETF Performance}

The immediate positive response of crypto-related ETFs following the GENIUS Act's passage provides clear market validation of regulatory clarity's importance \cite{chauhan2025crypto}. Institutional investors, who had previously remained cautious due to regulatory uncertainty, demonstrated renewed confidence through significant capital commitments.

This institutional engagement through regulated financial products represents a fundamental shift in how professional investors view digital assets. Rather than speculative investments suitable only for risk capital, stablecoins are increasingly viewed as legitimate infrastructure investments that can provide portfolio diversification and efficiency benefits.

The ETF performance also demonstrates the pent-up institutional demand for regulated cryptocurrency exposure \cite{chauhan2025crypto}. As additional regulatory clarity emerges and more sophisticated investment products become available, this demand is likely to drive continued growth in stablecoin adoption and integration.

\section{Challenges and Risks}

Table~\ref{tab:risk_mitigation} systematically analyzes the primary challenges facing stablecoin adoption and outlines how Banking 2.0 infrastructure addresses these risks through innovative technological and regulatory solutions.

\begin{table}[htbp]
    \caption{Addressing the Challenges: Risks and Mitigations}
    \label{tab:risk_mitigation}
    \centering
    \small
    \begin{tabular}{|p{2cm}|p{2.6cm}|p{2.6cm}|}
    \hline
    \textbf{The Challenge (Risk)} & \textbf{The Root Cause} & \textbf{The Banking 2.0 Solution (Mitigation)} \\
    \hline
    Peg Volatility (e.g., USDC depeg) & Concentrated, fiat-based reserves & Diversified, infrastructure-backed reserves \\
    \hline
    Regulatory Fragmentation & Inconsistent global laws & Harmonized frameworks (e.g., GENIUS Act as a model) \\
    \hline
    Technical Vulnerabilities (e.g., Quantum Threat) & Outdated cryptographic standards & Post-Quantum Cryptography (PQC) integration \\
    \hline
    \end{tabular}
    \begin{flushleft}
    \small
    \textit{Risk Mitigation Framework: Banking 2.0 infrastructure addresses systemic risks through diversification, regulatory harmonization, and advanced cryptographic standards, creating a more resilient financial ecosystem.}
    \end{flushleft}
\end{table}

\subsection{Volatility and Pegging Vulnerabilities}

Despite their name and design intentions, stablecoins can exhibit significant volatility when their pegging mechanisms fail or when they are tied to unstable underlying assets. The depegging of USDC in March 2023, triggered by exposure to Silicon Valley Bank's collapse, illustrated the risks inherent in reserve concentration and inadequate diversification \cite{spglobal2023stablecoins}.

This incident revealed that stablecoins pegged to traditional banking systems inherit the systemic risks of those systems. When USDC's reserves, held partially at Silicon Valley Bank, became uncertain due to the bank's failure, the stablecoin experienced significant price volatility and temporary loss of its dollar peg \cite{spglobal2023stablecoins}. This event demonstrated that the perceived stability of fiat-backed stablecoins can be illusory if the backing infrastructure is flawed.

The broader implication is that any stablecoin tied directly to government-issued currencies may be undermined by the same inflationary pressures, political manipulation, and systemic risks that affect traditional monetary systems \cite{fedreserve1971}, \cite{cea2025economic}. True stability requires more robust and diversified backing mechanisms that don't rely solely on the credibility and stability of individual institutions or governments.

\subsection{Regulatory Fragmentation and Implementation Challenges}

While landmark legislation like the GENIUS Act provides crucial clarity in the United States, the global regulatory landscape for stablecoins remains fragmented and inconsistent \cite{fsb2023framework}. This fragmentation creates operational challenges for stablecoin issuers and users, potentially limiting their effectiveness as global transaction media.

Different jurisdictions have adopted varying approaches to stablecoin regulation, creating compliance burdens that can limit innovation and increase costs. Some regions have embraced stablecoins with clear regulatory frameworks, while others have imposed restrictions or outright bans that fragment the global market \cite{fsb2023framework}.

The regulatory uncertainty also exposes stablecoin systems to political risk. Regulatory changes or hostile government actions can significantly impact stablecoin viability and adoption, creating volatility that undermines their primary value proposition of stability.

Additionally, the nascent nature of many regulatory frameworks means they may not adequately address all potential risks or operational scenarios. As stablecoin usage scales, previously unanticipated challenges may emerge that require regulatory adaptation, creating ongoing uncertainty for market participants.

\subsection{Technical and Operational Vulnerabilities}

Despite their generally strong security records \cite{benos2019economics}, stablecoins face ongoing technical challenges that could impact their stability and adoption. Smart contract vulnerabilities, oracle failures, and blockchain network congestion can all affect stablecoin performance and user confidence.

The complexity of modern stablecoin systems, particularly those that use algorithmic mechanisms or complex reserve structures, creates multiple potential failure points. While this complexity enables sophisticated functionality, it also increases the risk of unexpected interactions or cascading failures that could destabilize the system.

Scalability remains a challenge for many blockchain networks that support stablecoins. During periods of high network utilization, transaction fees can spike and confirmation times can increase, reducing the efficiency advantages that make stablecoins attractive compared to traditional payment systems \cite{bis2018cbdc}.

\subsection{Market Manipulation and Systemic Risks}

The relatively nascent and concentrated nature of stablecoin markets makes them potentially vulnerable to manipulation by large holders or coordinated attacks. While the transparency of blockchain systems makes some forms of manipulation more difficult, it also makes market positions and flows more visible to potential manipulators.

The concentration of stablecoin reserves in specific asset classes or institutions creates systemic risks that could affect the entire stablecoin ecosystem, as seen with the USDC depegging event \cite{spglobal2023stablecoins}. If major reserve holdings become impaired or if reserve custodians experience difficulties, multiple stablecoins could be affected simultaneously.

The interconnected nature of DeFi protocols that utilize stablecoins also creates potential contagion risks. Problems with one major stablecoin could cascade through various lending, trading, and yield-generating protocols, potentially amplifying the impact of any individual system failure.

\section{Future Outlook and Transformative Potential}

\subsection{The Hypothetical Ideal: Infrastructure-Backed Global Currency}

The ultimate vision for stablecoin evolution involves the development of internationally utilized digital currencies whose value derives not from the backing of individual national economies \cite{fedreserve1971}, but from robust global infrastructure or diversified baskets of stable assets \cite{renteria2021bitcoin}, \cite{chainalysis2023bitcoin}. Such stablecoins would significantly reduce reliance on the fluctuating fortunes and policy decisions of individual nations, fostering greater global financial resilience and stability.

This infrastructure-backed model could draw value from productive assets such as renewable energy infrastructure, transportation networks, telecommunications systems, and other essential services that provide real economic value regardless of political boundaries. By anchoring stablecoin value to productive capacity rather than government promises, such systems could provide more reliable stores of value over extended periods.

The diversification benefits of this approach extend beyond risk reduction to include the potential for value appreciation as global infrastructure expands and improves. Unlike fiat currencies that lose value through inflation \cite{cea2025economic}, infrastructure-backed stablecoins could appreciate as the underlying productive capacity grows, creating positive incentives for long-term saving and investment.

\subsection{Predicted Market Evolution and Growth Trajectories}

We predict an explosion in stablecoin sector growth, driven by the expansion of innovative networks like the Global Dollar Network \cite{gdn2025global} and continued development by established players like Paxos \cite{mcclain2024introducing}. Furthermore, the integration of stablecoins with artificial intelligence promises to create "smarter" banking systems, capable of more efficient risk management, automated compliance, and personalized financial services. AI-powered stablecoin systems could automatically adjust reserve compositions, optimize transaction routing, and provide predictive analytics that enhance stability and performance.

Particularly intriguing is the potential for stablecoins to play a role in addressing national debt burdens. President Trump's stablecoin reserve policy suggests that leveraging valuation gains from sovereign wealth funds that include cryptocurrencies could provide novel mechanisms for debt reduction \cite{labonte2025banking}. If successful, this approach could be adopted by other nations, creating a global trend toward strategic digital asset accumulation.

\subsection{Future-Proofing Through Technology and Regulation}

Ensuring the long-term viability and trustworthiness of stablecoins requires proactive measures in both legislative frameworks and technological advancement. Future legislation must be adaptive and comprehensive, providing clear guidelines for issuance, reserves, and interoperability while fostering innovation rather than constraining it.

The regulatory evolution must also address international coordination challenges. As stablecoins become more prominent in global commerce, the need for harmonized international standards becomes critical. Regulatory fragmentation that creates compliance burdens or limits interoperability could significantly hamper stablecoin adoption and effectiveness \cite{fsb2023framework}.

Technological future-proofing requires addressing emerging threats, particularly from quantum computing. The advent of practical quantum computers poses significant challenges to current cryptographic standards that underpin blockchain security. Integrating Post-Quantum Cryptography (PQC) into stablecoin infrastructure is essential for protecting these systems against future quantum attacks \cite{nist2022quantum}.

Companies like BTQ are developing quantum-secure technologies that can protect sensitive data and transactions from quantum computational threats \cite{nist2022quantum}. Embracing such PQC solutions ensures that stablecoins remain secure and resilient, safeguarding the integrity of Banking 2.0 infrastructure for decades to come.

\subsection{Supporting Evidence}

The establishment of President Trump's crypto advisory council, reportedly including key industry figures such as Ripple's CEO \cite{whitehouse2025digital}, demonstrates a high-level commitment to integrating digital assets into national financial strategy. Moreover, the increasing trend of sovereign wealth funds incorporating cryptocurrencies into their reserves lends strong credence to the future importance and acceptance of stablecoins at a national and international scale \cite{renteria2021bitcoin}, \cite{chainalysis2023bitcoin}.

\subsection{Broader Economic and Social Implications}

The widespread adoption of stablecoins promises a fundamental replatforming of global finance with profound implications extending far beyond payment efficiency. Enhanced financial inclusion for unbanked and underbanked populations could provide billions of people with access to stable stores of value and efficient payment systems for the first time \cite{sergio2025global}.

The productivity gains from reduced transaction friction could enable new forms of economic organization and collaboration. Micro-transactions become economically viable, enabling new business models for content, services, and digital goods. Cross-border commerce becomes more accessible to small businesses and individuals, democratizing access to global markets.

The competitive pressure from stablecoin efficiency could force traditional financial institutions to reduce costs and improve services, benefiting all users regardless of whether they directly adopt cryptocurrency technologies. This competitive dynamic could drive broader innovation and efficiency improvements throughout the financial system.

\section{Conclusion}

\subsection{Evidence Summary}

The convergence of technological maturation, regulatory clarity, and institutional adoption has created an unprecedented opportunity for stablecoins to reshape global banking infrastructure. More than a decade of real-world testing has demonstrated their superior fraud resistance \cite{benos2019economics} and efficiency compared to traditional systems \cite{bis2018cbdc}. Legislative developments, particularly the GENIUS Act of 2025 \cite{whitehouse2025genius}, \cite{housefinance2025genius}, have provided a regulatory foundation for mainstream adoption, while corporate pivots by major institutions like JPMorgan \cite{smith2025jpmorgan} and PayPal demonstrate growing recognition of stablecoins' competitive advantages.

The economic logic supporting stablecoin adoption is compelling. Traditional fiat systems face structural challenges including the inflation-productivity imbalance \cite{cea2025economic}, centralized vulnerabilities, and inefficient intermediary networks. Stablecoins address these challenges through technological innovation, creating sustainable competitive advantages that benefit end users through lower costs, faster settlements, and enhanced security. The evidence confirms our thesis: stablecoins have emerged as the inevitable foundation for next-generation banking infrastructure.

\subsection{Strategic Implications and Timeline}

Banking 2.0 represents a fundamental reimagining of how value is stored, transferred, and managed in the modern economy. Organizations should begin pilot programs within 12-18 months, governments should prioritize enabling regulatory frameworks, and businesses should evaluate stablecoin integration for cost reduction and market expansion.

Early adopters will benefit from competitive advantages in efficiency, innovation, and global market access. The transformation timeline is accelerating--institutional adoption is expected to reach critical mass within 24-36 months, driven by regulatory clarity and demonstrated economic benefits. The question is not whether stablecoins will reshape banking, but how quickly organizations will position themselves to benefit from this historic shift toward Banking 2.0.

\end{document}